\documentclass[
aps,
prb,
amsmath,
amssymb,
superscriptaddress,
reprint,
]{revtex4-2}

\usepackage{graphicx}   
\usepackage{dcolumn}    

\usepackage{siunitx}
\DeclareSIUnit\angstrom{\text {Å}}
\usepackage{mathtools}
\usepackage{bbold}
\usepackage{braket}
\usepackage[version=4]{mhchem}
\usepackage{bm}

\usepackage{hyperref}
\usepackage{booktabs}

\usepackage{xcolor}

\begin{document}

\title{A plane-wave auxiliary-field quantum Monte Carlo study of silicon point defects}

\author{Moritz Humer}
\thanks{Contact author: moritz.humer@vasp.at}
\affiliation{VASP Software GmbH, Berggasse 21/14, A-1090 Vienna, Austria}
\affiliation{Vienna Doctoral School in Physics, University of Vienna,  Kolingasse 14-16, A-1090 Vienna, Austria}
\author{Faruk Salihbegovic}
\affiliation{Computational Materials Physics, University of Vienna, Kolingasse 14-16, A-1090 Vienna, Austria}
\author{Martin Schlipf}
\affiliation{VASP Software GmbH, Berggasse 21/14, A-1090 Vienna, Austria}
\author{Zoran Sukurma}
\affiliation{Institute for Theoretical Physics, TU Wien, Wiedner Hauptstrasse 8-10/E136, A-1040 Vienna, Austria}
\author{Georg Kresse}
\affiliation{VASP Software GmbH, Berggasse 21/14, A-1090 Vienna, Austria}
\affiliation{Computational Materials Physics, University of Vienna, Kolingasse 14-16, A-1090 Vienna, Austria}

\date{\today}

\begin{abstract}
Accurately predicting the energetics of point defects remains a formidable challenge in computational materials science. For the tetrahedral silicon self-interstitial, for example, previous periodic coupled-cluster [CCSD(T)] and diffusion Monte Carlo (DMC) calculations disagree by about \SI{1.2}{\electronvolt}. Here, we present formation energies of silicon self-interstitials and vacancies using auxiliary-field quantum Monte Carlo (AFQMC) in a plane-wave basis formulated within the projector-augmented wave framework, and compare them with the random-phase approximation (RPA) and CCSD(T).
Using 16- and 64-atom cells, we demonstrate that an RPA-based correction effectively removes the two-body finite-size errors that remain after twist averaging in AFQMC for bulk silicon and for the dumbbell and hexagonal self-interstitials. With this correction, we obtain formation energies of \SI{4.4}{\electronvolt} and \SI{4.6}{\electronvolt} for these two self-interstitials in 16-atom cells, in good agreement with DMC and transcorrelated coupled-cluster results. For the tetrahedral interstitial and the Jahn--Teller-distorted vacancy, 16-atom cells are too small to obtain realistic formation energies; in 64-atom cells, we obtain \SI{5.7}{\electronvolt} and \SI{5.2}{\electronvolt}, respectively.
With identical setups in 16-atom cells, CCSD(T) yields systematically lower formation energies than AFQMC, with the largest difference, \SI{250}{\milli\electronvolt}, for the tetrahedral interstitial. Our results indicate that the earlier overestimation of CCSD(T) for the tetrahedral interstitial stems from the finite-size treatment in the coupled-cluster workflow rather than from multi-configurational character of the defect.
\end{abstract}

\maketitle

\section{Introduction}\label{Sec.I_introduction}

The electrical, optical, and catalytic properties of materials are often significantly altered by the presence of point defects, which introduce defect states into the band gap~\cite{Li2020Defect,Pastor2022Electronic,Mosquera-Lois2023Imperfections,Tan2026Multifunctional}. Consequently, as the semiconductor industry advances toward the single-nanometer scale, the precise manipulation and understanding of vacancies and self-interstitials --- particularly in silicon --- are crucial. Experimental defect identification, however, is difficult as it typically requires a combination of complementary techniques~\cite{Alkauskas2016Tutorial}.

Modeling has thus become indispensable for studying point defects, yet quantitatively predicting their energetics remains a formidable challenge. Density functional theory (DFT)~\cite{kohn1965self}, with its unsurpassed balance between accuracy and computational efficiency, has been used extensively to describe defects in semiconductors and insulators~\cite{Van1987Theoretical,Van2006Hydrogen,estreicher2006theory,Freysoldt2014First}. However, within the local-density approximation (LDA) and the generalized gradient approximation (GGA), DFT severely underestimates band gaps and often fails to describe charge localization arising from narrow bands or local lattice distortions near defects. These failures are mainly attributed to the inherent self-interaction error in local and semilocal functionals, which causes defect formation energies to be underestimated relative to experiment~\cite{Schultz2006Theory,Rinke2009Defect,Stewart2020Reexamining}. Hybrid functionals reduce the self-interaction error by including a fraction of exact exchange, which largely mitigates the band-gap problem; however, their empirical parametrization necessitates \textit{ab initio} many-body benchmarks. Moreover, density functionals are not, in general, systematically improvable, and it often remains \textit{a priori} unclear which exchange-correlation functional is best suited for a given problem, as different functionals can yield qualitatively or at least quantitatively different results~\cite{montero2024,bazantes2026rationalizing}.

To overcome these limitations, researchers have turned to many-body methods such as the random-phase approximation (RPA)~\cite{nozieres1958correlation,langreth1977exchange,miyake2002total,fuchs2005describing,furche2008developing}, coupled-cluster (CC) theory~\cite{vcivzek1966correlation,vcivzek1971correlation}, and stochastic methods like diffusion Monte Carlo (DMC)~\cite{anderson1976quantum,ceperley1977monte}. However, the RPA systematically underestimates formation energies relative to DMC results~\cite{kaltak2014cubic,Leung1999Calculations,Batista2006Comparison,Parker2011Accuracy}. Moreover, for the silicon self-interstitial landscape, a substantial discrepancy has emerged between DMC and coupled cluster including singles, doubles, and perturbative triples [CCSD(T)]: the CCSD(T) formation energy of the tetrahedral interstitial exceeds the DMC value by approximately \SI{1.2}{\electronvolt}~\cite{Salihbegovi_Fomration,Parker2011Accuracy}. The authors of Ref.~\onlinecite{Salihbegovi_Fomration} attribute this discrepancy to strong correlation and multi-configurational character of defect states in the gap, which challenge single-reference CC methods. This discrepancy highlights the difficulty of predicting the defect energetics accurately and motivates additional studies using other high-level many-body methods.

Over the past decade, auxiliary-field quantum Monte Carlo (AFQMC) has emerged as a promising stochastic many-body method, offering near-chemical accuracy at low-order polynomial cost. It is highly parallelizable and, when formulated in a plane-wave (PW) basis, scales cubically with system size.
AFQMC also has a distinct advantage over other quantum Monte Carlo (QMC) methods: whereas DMC~\cite{anderson1976quantum,ceperley1977monte,foulkes2001quantum,nemec2010benchmark,hammond1987valence} requires the locality approximation to use non-local pseudopotentials~\cite{mitavs1991nonlocal,casula2006beyond,casula2010size,zen2019new,anderson2021nonlocal}, AFQMC requires no such approximation~\cite{suewattana2007phaseless}. Consequently, it can be readily formulated within the projector-augmented wave (PAW)~\cite{bloechl1994} framework. The PAW method achieves near-all-electron accuracy while using smooth pseudo orbitals, thereby reducing the required PW cutoff compared to norm-conserving pseudopotentials. Furthermore, it offers better transferability than pseudopotentials suited for DMC, particularly for heavier elements and transition metals.

Historically, AFQMC was formulated as a path-integral method~\cite{hirsch1985two,sugiyama1986auxiliary}. Because this approach was severely limited by the fermionic sign problem~\cite{loh1990sign,zhang1995constrained}, \citeauthor{zhang1995constrained} reformulated AFQMC as an open-ended random walk in the space of non-orthogonal Slater determinants~\cite{zhang1995constrained}. This advance led to constrained-path AFQMC, which has proven highly successful for model Hamiltonians. Extending the method to realistic many-body Hamiltonians, however, transforms the sign problem into a more general phase problem. To address this challenge, \citeauthor{zhang2003quantum} introduced the phaseless approximation~\cite{zhang2003quantum}. PW AFQMC implementations~\cite{suewattana2007phaseless} have since been used to study phase transitions~\cite{purwanto2009pressure}, excited states~\cite{ma2013excited}, the uniform electron gas~\cite{lee2019auxiliary}, and to obtain benchmark charge densities~\cite{chen2021ab}. Recent extensions include forces and stresses in solids~\cite{chen2023computation}.

In a previous work~\cite{Humer2026Auxiliary}, we presented a PW AFQMC implementation within the PAW framework~\cite{bloechl1994} in the Vienna \textit{ab initio} Simulation Package (VASP)~\cite{kresse1996efficient,kresse1999}. This implementation offers cubic scaling in time and quadratic scaling in memory. By incorporating the exact inverse of the PAW overlap operator in the imaginary-time propagation~\cite{Walker2007Ultrasoft,Humer2026Auxiliary}, it naturally includes all excitations within the PW cutoff sphere. Convergence is thus governed solely by the PW cutoff.
We benchmarked this implementation on the lattice constants of typical bulk semiconductors and insulators. By embedding AFQMC in second-order M{\o}ller--Plesset (MP2) and RPA calculations, we showed that AFQMC recovers both the long-range screening missing in MP2 and the higher-order exchange missing in the RPA~\cite{Humer2026Auxiliary}. Furthermore, we established that an RPA reference converges fastest with supercell size. The AFQMC lattice constants systematically improved upon MP2 and RPA, matched or exceeded the accuracy of DMC, and agreed with periodic CCSD(T) for diamond. Having established this approach's accuracy for bulk structural properties, we now apply it to the formation energies of silicon self-interstitials and vacancies.

The remainder of the paper is structured as follows: Sec.~\ref{Sec.II_theory} provides an overview of the AFQMC method formulated in a PW basis, Sec.~\ref{Sec.III_computational_details} describes the workflow and the computational details, followed by the results in Sec.~\ref{Sec.IV_results}.

\section{Theory}\label{Sec.II_theory}
This section provides a brief overview of the AFQMC method formulated in a PW basis; a self-contained description within the PAW framework is given in Ref.~\onlinecite{Humer2026Auxiliary}. We use atomic units throughout.

AFQMC obtains the ground state $\ket{\Phi_0}$ of an interacting many-body system via projection in imaginary time
\begin{equation}
    \ket{\Phi_0} = \lim_{n\to\infty} \left[ e^{-\tau(\hat{H}-E_0)}\right]^n\ket{\Psi_\mathrm{I}},
    \label{Eq.II.1_projection}
\end{equation}
where the initial wavefunction $\ket{\Psi_\mathrm{I}}$ must have non-zero overlap with the ground state, \textit{i.e.}, $\braket{\Phi_0 | \Psi_\mathrm{I}} \neq 0$. Because the ground-state energy $E_0$ is generally unknown it requires approximation, \textit{e.g.}, from an underlying mean-field calculation.

Within the Born--Oppenheimer approximation, the Hamiltonian is naturally decomposed into a zero-body contribution $\hat{H}_0$, describing the ion--ion interactions; a one-body contribution $\hat{H}_1$, containing the electronic kinetic energy and the electron--ion interaction; and a two-body contribution $\hat{H}_2$, describing the electron--electron interaction. The short-time propagator is then factorized according to these contributions using, \textit{e.g.}, a second-order Trotter--Suzuki decomposition~\cite{trotter_1959,suzuki_1976}
\begin{equation}
    e^{-\tau(\hat{H}-E_0)} \approx e^{-\tau(\hat{H}_0-E_0)} e^{-\frac{\tau}{2}\hat{H}_1} e^{-\tau\hat{H}_2} e^{-\frac{\tau}{2}\hat{H}_1}.
    \label{Eq.II.1_trotter_suzuki}
\end{equation}

Next, the two-body operator $e^{-\tau\hat{H}_2}$ is rewritten in terms of one-body operators via a Hubbard--Stratonovich (HS) transformation~\cite{Hubbard_1959,stratonovich_1957}. This, however, requires the two-body Hamiltonian to be expressed in a quadratic form. In the case of a PW basis, $\hat{H}_2$, given in second quantization, is rewritten in terms of density--density interactions by applying the fermionic anticommutation relations:
\begin{equation}
    \hat{H}_2 = \frac{1}{2} \sum_{\bm{G}\neq 0} \frac{4\pi}{\Omega \bm{G}^2}\, \hat{n}^\dagger(\bm{G})\, \hat{n}(\bm{G}) + \hat{H}_\mathrm{self}.
    \label{Eq.II.1_density_hamiltonian}
\end{equation}
Here, $\hat{n}(\bm{G})$ denotes the Fourier components of the density operator, $\bm{G}$ is a reciprocal lattice vector, and $\Omega$ is the system's volume. The divergent $\bm{G} = 0$ component is omitted because it cancels against the $\bm{G} = 0$ components of the ion--ion and electron--ion interactions~\cite{Ihm_1979}. The additional term $\hat{H}_\mathrm{self}$ corrects for the self-interactions introduced by the density--density interactions in the first term of Eq.~\eqref{Eq.II.1_density_hamiltonian}. Since $\hat{H}_\mathrm{self}$ is a one-body operator, it is absorbed into the one-body Hamiltonian: $\hat{H}_1 \leftarrow \hat{H}_1 + \hat{H}_\mathrm{self}$.

To improve the efficiency of the Monte Carlo (MC) sampling, we reduce the magnitude of the two-body term by subtracting the mean-field background from the density operators~\cite{baer1998shifted,suewattana2007phaseless}. For brevity, this shift is suppressed in the notation; explicit expressions are given in Ref.~\onlinecite{Humer2026Auxiliary}. The HS transformation is then applied to the exponential of the first term in Eq.~\eqref{Eq.II.1_density_hamiltonian}
\begin{equation}
    \begin{split}
        e^{-\tau\hat{H}_2}
         \approx \prod_{\bm{G}\neq0} \int \mathrm{d}^2\sigma_{\bm{G}}~ g(\sigma_{\bm{G}}) e^{ -\tau \sigma^*_{\bm{G}} \hat{n}(\bm{G})},
    \end{split}
    \label{Eq.II.1_HS_transformation}
\end{equation}
where $g(\sigma_{\bm{G}})$ denotes a complex normal distribution with variance $\frac{4\pi}{\tau\Omega\bm{G}^2}$. Because the density operator is real-valued in real space, the auxiliary fields $\sigma_{\bm{G}}$ must satisfy $\sigma^*_{\bm{G}} = - \sigma_{-\bm{G}}$. The transformation in Eq.~\eqref{Eq.II.1_HS_transformation} is exact for pairwise commuting operators; however, for general operators, the leading error is $\mathcal{O}(\tau^2)$. The HS transformation replaces an interacting system by a non-interacting system coupled to fluctuating auxiliary fields $\bm{\sigma}$. The full two-body interaction is recovered by the multidimensional integral over these fields.

To evaluate the multidimensional integral in Eq.~\eqref{Eq.II.1_HS_transformation} efficiently, one introduces a walker ensemble, sampling the integral by a single auxiliary-field configuration $\bm{\sigma}_k^w$ for each walker $w$ and time step $k$. Combining Eqs.~\eqref{Eq.II.1_HS_transformation} and~\eqref{Eq.II.1_trotter_suzuki} for a given field configuration yields the stochastic short-time propagator $P(\bm{\sigma})$, a product of exponentials of one-body operators. Each walker in the ensemble is described by a single Slater determinant $\ket{\Psi_k^w}$ and a weight $W_k^w$, which is in general complex. At time step $k = 0$, the walker ensemble is initialized as
\begin{equation}
    \ket{\Psi_0^w} = \ket{\Psi_\mathrm{I}};\qquad W_0^w = \braket{\Psi_\mathrm{T} |\Psi_\mathrm{I}},
    \label{Eq.II.1_walker_initial}
\end{equation}
where $\ket{\Psi_\mathrm{I}}$ is a single Slater determinant --- usually taken from a mean-field calculation --- and $\ket{\Psi_\mathrm{T}}$ denotes the trial wavefunction, a best guess of the many-body ground state that guides the random walk. To improve the sampling efficiency, we employ the importance-sampling transformation introduced by \citeauthor{zhang2003quantum} together with the force bias $\bar{\bm{\sigma}}$~\cite{zhang2003quantum}. This transformation introduces an importance function $I(\bm{\sigma},\bar{\bm{\sigma}},\Psi)$ that depends on the trial wavefunction and the force bias. Explicit expressions are given in Ref.~\onlinecite{Humer2026Auxiliary}. Within the phaseless approximation~\cite{zhang2003quantum}, the walkers and their weights are then updated as
\begin{align}
    \ket{\Psi_{k+1}^w} &= P(\bm{\sigma}_k^w - \bar{\bm{\sigma}}_k^w)\ket{\Psi_{k}^w}, \label{Eq.II.1_eom_walker} \\
    W_{k+1}^w &= \left| I(\bm{\sigma}_k^w,\bm{\bar\sigma}_k^w,\Psi_k^w) \right|  W_{k}^w \max \left (0,\cos(\Delta\theta_k^w) \right), \label{Eq.II.1_eom_weight}
\end{align}
where $\Delta\theta_k^w = \Im \log \frac{\braket{\Psi_\mathrm{T}|\Psi_{k+1}^w}}{\braket{\Psi_\mathrm{T}|\Psi_k^w}}$ denotes the phase change in the walker weight from one step to the next. The phaseless approximation ensures that the walker weights remain real and non-negative. Because the initial wavefunction $\ket{\Psi_\mathrm{I}}$ is a single Slater determinant and the stochastic short-time propagator $P(\bm{\sigma})$ is a product of exponentials of one-body operators, Thouless' theorem guarantees that the walkers remain single Slater determinants throughout their trajectory. This holds only because the integral in Eq.~\eqref{Eq.II.1_HS_transformation} is sampled by exactly one auxiliary-field configuration per walker and time step.

Finally, after an equilibration phase during which excited states in the initial wavefunction decay exponentially, the ground-state energy is estimated by the time-averaged estimator:
\begin{equation}
    \bar{E} = \frac{\sum_{kw} W_k^w E_\mathrm{loc}(\Psi_k^w)}{\sum_{kw} W_k^w},
    \label{Eq.IV.2.1_time_average_energy}
\end{equation}
where the local energy $E_\mathrm{loc}$ is given by the mixed estimator:
\begin{equation}
    E_\mathrm{loc}(\Psi_k^w) = \frac{\braket{\Psi_\mathrm{T}  | \hat{H} | \Psi_k^w}}{\braket{\Psi_\mathrm{T} | \Psi_k^w}}.
    \label{Eq.IV.2.1_local_energy}
\end{equation}
The mixed estimator is exact only for observables that commute with the propagator and hence with the Hamiltonian. The local energy is evaluated using the generalized Wick theorem~\cite{Balian_Nonunitary}, splitting Eq.~\eqref{Eq.IV.2.1_local_energy} into a one-body, a Hartree-like, and an exchange-like contribution. A comprehensive discussion of the local energy evaluation within the PAW method has been given in Ref.~\onlinecite{Humer2026Auxiliary}.

\section{Computational details}\label{Sec.III_computational_details}

All AFQMC, RPA, and HSE calculations were performed using VASP~\cite{kresse1996efficient,kresse1999} within the PAW framework~\cite{bloechl1994} in a PW basis. The CCSD(T) calculations were performed using the open-source code coupled cluster for solids (cc4s)~\cite{Hummel2017Low,liao2016communication,irmler2021focal}, which is interfaced with VASP. Since this work employs correlated-wavefunction methods, we use the \texttt{Si\_GW} PAW potential. This potential is designed to reproduce accurate scattering properties well above the vacuum level and has been shown to perform reliably in many-body calculations. Table~\ref{Tab.III_pseudopotentials} summarizes the parameters of the employed PAW potential.

\begin{table}
\caption{\label{Tab.III_pseudopotentials} PAW potential used in this work. Valence orbitals and VASP database labels are listed. Radial cutoffs are specified as $n\times r_\mathrm{cut}$, where $n$ is the number of projectors and $r_\mathrm{cut}$ is the cutoff radius (in atomic units) for each angular momentum quantum number channel. The local potential matches the all-electron potential outside $r_\mathrm{core}$. The SHA-256 of the respective potentials are given in the table footnotes.}
\begin{ruledtabular}
\begin{tabular}{llllllll}
atom      & valence   & label & s             &  p            &   d           &   f          &$r_\mathrm{core}$ \\
\midrule
Si        & $3s^23p^2$    & GW\footnote{\scriptsize \texttt{9ee269d4247e69d62ab9db642b0c3f6e4981ddd1f39659ad08dec53e20e43eb1}}  & 2$\times$ 1.9 & 2$\times$ 1.9 & 2$\times$ 1.9 & 1$\times$1.9 &1.9 \\
\end{tabular}
\end{ruledtabular}
\end{table}

We considered 17- and 65-atom silicon self-interstitial cells and 15- and 63-atom vacancy cells. The associated bulk supercells contained 16 and 64 atoms, respectively. The geometries and atomic positions were adopted from Ref.~\onlinecite{kaltak2014cubic}, where they were relaxed with DFT using the Perdew--Burke--Ernzerhof (PBE)~\cite{perdew1996generalized} functional at fixed cell shapes and volumes. In the following, ``16-atom cells'' collectively refers to the smaller bulk and defect supercells ($16\pm1$ atoms). Similarly, ``64-atom cells'' refers to the larger systems ($64\pm1$ atoms). The geometries are provided in the supplementary material.

We considered three silicon self-interstitial configurations. The most stable is the ``dumbbell'' (X), in which two Si atoms aligned along the $\left[ 110 \right]$ direction symmetrically share one lattice site. Second in stability is the ``hexagonal hollow'' (H), in which the additional Si atom sits equidistant from six lattice atoms forming a hexagonal ring. Within PBE, this configuration is unstable: the central atom moves orthogonally out of the hexagonal plane. We did not consider this distorted configuration, since previous calculations found the associated energy changes to be well below the statistical uncertainties we have been able to achieve in the present work~\cite{kaltak2014cubic}. The least stable is the ``tetrahedral'' (T) interstitial, in which the additional Si atom is tetrahedrally coordinated to four lattice atoms. Its symmetry yields a threefold-degenerate highest occupied orbital that hosts two electrons, suggesting susceptibility to a Jahn--Teller distortion. However, \citeauthor{kaltak2014cubic} report that T remains stable within PBE for a 64-atom cell~\cite{kaltak2014cubic}.

The vacancy (V) is created by removing one Si atom from the bulk. Like the T interstitial, it features a threefold-degenerate highest occupied orbital with two electrons and is well known to undergo Jahn--Teller distortion~\cite{Corsetti2011System}. For the 16-atom cells, we studied the undistorted vacancy; for the 64-atom cell, we studied the Jahn--Teller-distorted vacancy (VJT), since it is more stable in DFT.

\subsection{AFQMC calculations}\label{Sec.III.1_afqmc_calculations}
All AFQMC calculations were performed using the phaseless approximation~\cite{zhang2003quantum} with a single-determinant Hartree--Fock (HF) trial and initial wavefunction. Importantly, the underlying mean-field HF calculations always yielded gapped systems. We employed 2560 walkers per k-point throughout this work. Formation energies for the 16-atom cells were extrapolated to the zero-time-step limit using four imaginary-time steps $\tau$ (see Sec.~\ref{Sec.IV.1_time-step_errors}). For the 64-atom cells, such an extrapolation was computationally prohibitive, and a time step of $\tau=\SI{1.0e-3}{\per\electronvolt}$ was employed. We rebalanced the walker population roughly every \SI{0.025}{\per\electronvolt}. To alleviate some of the computational cost, we sampled the exchange contribution to the local energy only every \SI{0.01}{\per\electronvolt} (every 10 to 40 steps, depending on the time step). The statistical uncertainties were estimated via block averaging. We targeted statistical uncertainties in the AFQMC formation energies of approximately \SI{15}{\milli\electronvolt} for the 16-atom cells and \SIrange{100}{200}{\milli\electronvolt} for the 64-atom cells. Finally, we applied rare-event control to the local energy, the importance weight, and the force bias as described in Ref.~\onlinecite{sukurma2024toward}.

\begin{table}
\caption{\label{Tab.III.1_parameters}
Computational parameters for 16- and 64-atom cells: BZ sampling scheme, PW cutoff $E_\mathrm{cut}$, and auxiliary PW cutoff $E_\mathrm{cut}^\chi$ (all cutoffs in \si{\electronvolt}). The suffixes -conv and -ta denote converged and twist-averaged calculations. The latter use a generalized regular k-point grid equivalent to a $4\times4\times4$ Monkhorst--Pack grid for the conventional eight-atom cubic cell; the number of irreducible k-points is given in parentheses.}
\begin{ruledtabular}
\begin{tabular}{lccc}
Method & BZ sampling & $E_\mathrm{cut}$ & $E_\mathrm{cut}^\chi$\\
\midrule
16-atom cells & & &\\
HSE-conv   & $\Gamma$-centered $8\times8\times8$ & 350 & \\
RPA-conv   & $\Gamma$-centered $7\times7\times7$ & 450 & 300 \\
RPA-ta     & twist average (2--10 k-points) & 250 & 562 \\
AFQMC-ta   & twist average (2--10 k-points) & 250 & 562 \\
CCSD(T)-ta & twist average (2--10 k-points) & 250 & 562 \\
\midrule
64-atom cells & & &\\
HSE-conv   & $\Gamma$-centered $3\times3\times3$ & 350 &\\
RPA-conv   & $\Gamma$-centered $3\times3\times3$ & 350 & 233 \\
RPA-ta     & twist average (1--3 k-points) & 250 & 562 \\
AFQMC-ta   & twist average (1--3 k-points) & 250 & 562 \\
\end{tabular}
\end{ruledtabular}
\end{table}

Presently, our AFQMC implementation samples the Brillouin zone (BZ) at a single k-point. To reduce the resulting one-body finite-size errors, we employ twist averaging: independent calculations are performed at the irreducible k-points $\bm{k}_i$ of a generalized regular k-point grid, and the twist-averaged energy is obtained as
\begin{equation}
    E_\mathrm{ta} = \sum_{\bm{k}_i} w_{\bm{k}_i} E(\bm{k}_i),
    \label{Eq.III.1_twist_averaging}
\end{equation}
where $w_{\bm{k}_i}$ denotes the symmetry weight of $\bm{k}_i$ and the subscript ``ta'' denotes twist-averaged quantities. The number of MC samples assigned to each irreducible k-point was chosen proportional to its symmetry weight.

To ensure a constant k-point density across the 16- and 64-atom cells, the twist averaging is based on a generalized regular k-point grid corresponding to a $4\times4\times4$ Monkhorst--Pack grid of the conventional eight-atom cubic cell. Depending on the symmetry of each defect cell, this grid contains two to ten irreducible k-points for the 16-atom cells and one to three for the 64-atom cells.

Equation~\eqref{Eq.III.1_twist_averaging} is applied in the same way to the other correlated methods studied in this work. Importantly, we twist-average both the mean-field and correlation contributions and employ identical PW cutoffs for both. This ensures a consistent cancellation between exchange and correlation contributions.

Twist averaging efficiently removes one-body finite-size errors, originating from single-particle effects. Unlike explicit BZ sampling, however, it is restricted to zero momentum transfer ($\bm{q}=0$) within the supercell and therefore does not remove two-body finite-size errors, which stem from missing long-wavelength exchange and correlation contributions.

In our previous study of structural properties~\cite{Humer2026Auxiliary}, we demonstrated that, for sufficiently large supercells, AFQMC accurately captures both short- and long-range correlation effects. Short-range correlation effects, associated with the large-$\bm{q}$ limit, converge rapidly with supercell size. The remaining finite-size errors are dominated by the slowly converging long-range ($\bm{q}\rightarrow0$) exchange and correlation energy contributions. Since the RPA includes exact exchange and accurately recovers these long-range contributions, it provides a robust correction for the remaining two-body finite-size errors.
We thus proposed an RPA-based correction to the twist-averaged AFQMC total energy
\begin{equation}
    \Delta E^\mathrm{RPA} = E^\mathrm{RPA}_\mathrm{conv} - E^\mathrm{RPA}_\mathrm{ta},
    \label{Eq.III.1_rpa_correction}
\end{equation}
where the subscript ``conv'' denotes a calculation converged with respect to both the BZ sampling and the basis set (see Sec.~\ref{Sec.III.2_rpa_calculations}). Consequently, $\Delta E^\mathrm{RPA}$ corrects not only the two-body finite-size errors but also the basis-set incompleteness of the twist-averaged calculations, which use a lower PW cutoff (see Table~\ref{Tab.III.1_parameters}).
The corrected AFQMC+$\Delta$RPA total energy is then
\begin{equation}
    E^\mathrm{AFQMC+\Delta RPA} = E^\mathrm{AFQMC}_\mathrm{ta} + E^\mathrm{RPA}_\mathrm{conv} - E^\mathrm{RPA}_\mathrm{ta}.
    \label{Eq.III.1_embedding}
\end{equation}

Since the correction scheme relies on an energy difference, the twist-averaged RPA and AFQMC calculations employ identical computational parameters --- PW cutoff, auxiliary PW cutoff, treatment of the PAW one-center terms, and FFT grids --- summarized in Table~\ref{Tab.III.1_parameters}. The RPA calculations are described in Sec.~\ref{Sec.III.2_rpa_calculations}.

Unlike the RPA correlation energy, the AFQMC correlation energy cannot be extrapolated with respect to the auxiliary PW cutoff $E_\mathrm{cut}^{\chi}$, which is used to evaluate the Hartree and exchange energies. We therefore set $E_\mathrm{cut}^{\chi}$ equal to the kinetic energy of the largest reciprocal lattice vector of the FFT grid. This means, in practice, no PW cutoff is used, and all PWs are included in the fluctuation terms.

\subsection{RPA calculations}\label{Sec.III.2_rpa_calculations}
In this work, we provide converged and twist-averaged RPA formation energies and use the RPA to correct AFQMC. The converged RPA energies serve as the reference $E^\mathrm{RPA}_\mathrm{conv}$ in the RPA-based correction; the twist-averaged energies $E^\mathrm{RPA}_\mathrm{ta}$ are computed separately.

For the converged RPA calculations, we sampled the BZ with $\Gamma$-centered k-point grids. For the 16-atom cells, the RPA formation energies were converged with respect to both the k-point grid and the PW cutoff $E_\mathrm{cut}$. For the 64-atom cells, a full convergence study was too costly; we used the largest affordable setup, at which the computational parameters changed the formation energies by only \SIrange{20}{50}{\milli\electronvolt}. The RPA correlation energies were extrapolated to the infinite auxiliary basis-set limit using the built-in extrapolation scheme implemented in VASP~\cite{harl2008cohesive}. The computational parameters are summarized in Table~\ref{Tab.III.1_parameters}.

The twist-averaged RPA calculations use the same settings as the AFQMC calculations (see Sec.~\ref{Sec.III.1_afqmc_calculations} and Table~\ref{Tab.III.1_parameters}). In particular, $E_\mathrm{cut}^\chi$ is set to the kinetic energy of the largest reciprocal lattice vector of the FFT grid and no extrapolation to the infinite auxiliary basis-set limit is performed.

Owing to the high defect density, the small supercells may be metallic depending on the BZ sampling. For all RPA calculations, we therefore adopted the strategy proposed by \citeauthor{harl2010assessing}~\cite{harl2010assessing}, who found that evaluating the exact-exchange (EXX) and RPA correlation energies with identical BZ sampling gives the fastest k-point convergence for metallic systems. We also used identical PW cutoffs for the EXX and RPA correlation energies, and omitted the long-wavelength ($\bm{q}=0$) contributions. To treat partial occupancies near the Fermi level consistently, all RPA calculations were performed within the Matsubara framework~\cite{Kaltak2020Minimax} and the EXX energies were corrected~\cite{harl2010assessing}.
We used PBE orbitals as the single-particle orbital basis and included all unoccupied (virtual) orbitals determined by the PW cutoff in the RPA calculations.

For comparison with a hybrid functional, we also computed formation energies with the screened hybrid functional HSE06~\cite{Heyd2003Hybrid,Krukau2006Influence,Paier2006Screened}, denoted HSE below, using converged k-point grids and PW cutoffs (see Table~\ref{Tab.III.1_parameters}).

\subsection{CCSD(T) calculations}\label{Sec.III.3_ccsd(t)_calculations}
We performed the CCSD(T) calculations following the workflow described in Ref.~\onlinecite{Salihbegovi_Fomration} using the coupled-cluster code cc4s~\cite{Hummel2017Low,liao2016communication,irmler2021focal}. The code is interfaced with VASP, which is used to generate the reference HF wavefunction and the required intermediates. To ensure a consistent setup with our AFQMC calculations, we chose identical settings for these preparatory VASP calculations (see Table~\ref{Tab.III.1_parameters}). CCSD(T) calculations were performed for the 16-atom cells only.

Like our AFQMC implementation, cc4s samples the BZ at a single k-point; we thus use the same twist averaging over the irreducible k-points of a regular grid described in  Sec.~\ref{Sec.III.1_afqmc_calculations}. This is in contrast to Ref.~\onlinecite{Salihbegovi_Fomration}, where the authors averaged over ten random k-points.

The correlation energy converges slowly with the number of unoccupied orbitals. Unlike in our AFQMC and RPA calculations, including all unoccupied orbitals in CCSD(T) is not feasible because of the prohibitive scaling of the perturbative triples contribution. We therefore employed MP2 natural orbitals~\cite{Grueneis2011Natural}, which result in a faster convergence to the basis-set limit than canonical HF orbitals. We used 20 natural orbitals per occupied state, in line with Ref.~\onlinecite{Salihbegovi_Fomration}, which found that 10--20 natural orbitals per occupied state yield converged formation energies.
The residual basis-set incompleteness error introduced by this truncation is corrected by the pair-specific focal-point correction (FPC)~\cite{irmler2021focal} implemented in cc4s. With 20 natural orbitals per occupied state, the FPC lowers the formation energies by \SIrange{20}{50}{\milli\electronvolt}, depending on the defect.

Our CCSD(T) workflow differs from that of Ref.~\onlinecite{Salihbegovi_Fomration} in its treatment of finite-size errors. First, we twist-average both the HF and the correlation contributions, whereas Ref.~\onlinecite{Salihbegovi_Fomration} combined twist-averaged correlation energies with k-point-converged HF energies. Second, we do not apply the finite-size correction based on structure-factor interpolation~\cite{liao2016communication} that Ref.~\onlinecite{Salihbegovi_Fomration} used to correct the CCSD correlation energy for two-body finite-size errors. Both workflows use the FPC. Our twist-averaged CCSD(T) results are therefore directly comparable to the twist-averaged AFQMC results, whereas the CCSD(T) results of Ref.~\onlinecite{Salihbegovi_Fomration} are corrected toward the thermodynamic limit of the 16-atom cell.
For the comparison with the literature in Sec.~\ref{Sec.IV.4_literature_comparison}, we also apply the RPA-based correction to the twist-averaged CCSD(T) energies, denoted CCSD(T)+$\Delta$RPA. Since $\Delta E^\mathrm{RPA}$ does not depend on the correlated method, the correction shifts the CCSD(T) and AFQMC formation energies by identical amounts and leaves their difference unchanged.

\subsection{Computational cost}
AFQMC and CCSD(T) differ markedly in computational cost. For the 16-atom cells, the time-step-extrapolated AFQMC formation energies required approximately \SI{1.6E6}{core~hours} using AMD EPYC 7713 nodes, compared to approximately \SI{1.0E5}{core~hours} for CCSD(T). At first sight, AFQMC is therefore the more expensive method. This comparison, however, favors CCSD(T): AFQMC propagates the walkers in the full PW basis and thus includes all unoccupied orbitals within the cutoff sphere, whereas CCSD(T) is restricted to 20 natural orbitals per occupied state and relies on the FPC for the introduced basis-set incompleteness error.
Including all unoccupied orbitals in CCSD(T) would be prohibitive, because the perturbative triples scale as $N_\mathrm{o}^3 N_\mathrm{v}^4$ with the numbers of occupied ($N_\mathrm{o}$) and unoccupied ($N_\mathrm{v}$) orbitals. Likewise, the memory requirements of AFQMC scale quadratically with system size, whereas those of CCSD(T) scale quartically.
The same scaling limits the accessible cell size. At a fixed number of natural orbitals per occupied state, increasing the cell from 16 to 64 atoms raises the cost of the triples contribution by a factor $4^7$, compared to a factor $4^3$ per time step in AFQMC. The 64-atom AFQMC calculations required approximately \SI{2.0E6}{core~hours}, whereas the corresponding CCSD(T) calculations were computationally infeasible.

\begin{figure*}[!ht]
    \centering
    \includegraphics{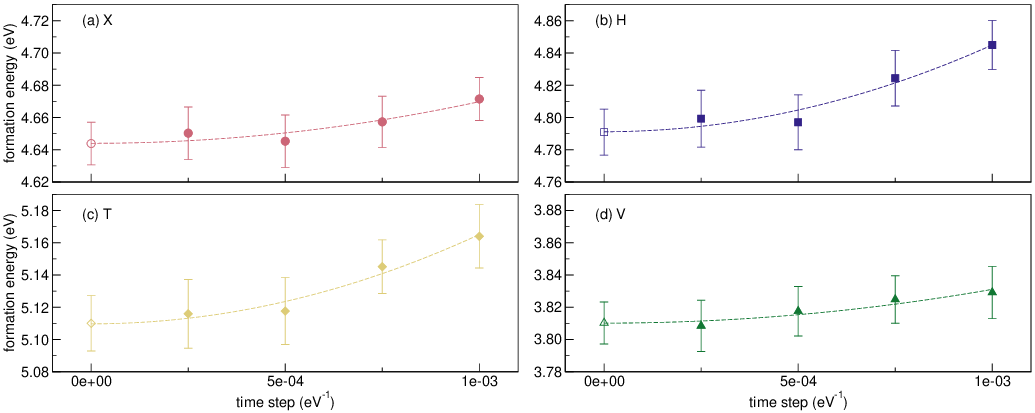}
    \caption{Time-step dependence of twist-averaged AFQMC formation energies (in \si{\electronvolt}) for 16-atom silicon self-interstitials [(a) X, (b) H, (c) T] and the vacancy [(d) V]. Dashed lines show quadratic fits $E(\tau) = E_0 + \beta\tau^2$; open symbols at $\tau=0$ indicate extrapolated values. No RPA correction is applied. Error bars denote one standard deviation $(1\sigma)$.}
    \label{Fig.IV.1_time-step_error}
\end{figure*}

\section{Results}\label{Sec.IV_results}
In this section, we first discuss the impact of time-step errors in the AFQMC formation energies (Sec.~\ref{Sec.IV.1_time-step_errors}) and assess how well the RPA correction mitigates two-body finite-size errors (Sec.~\ref{Sec.IV.2_bulk_energies}). Then we present the HSE, RPA, AFQMC, and CCSD(T) formation energies of the self-interstitials and the vacancy in 16- and 64-atom cells (Sec.~\ref{Sec.IV.3_formation_energies}), and finally compare our results with the literature (Sec.~\ref{Sec.IV.4_literature_comparison}).

Given the total energy $E$ of a defect and the associated bulk system, the formation energy is defined as
\begin{equation}
    E_\mathrm{F} = E_\mathrm{defect} - \frac{N_\mathrm{defect}}{N_\mathrm{bulk}} E_\mathrm{bulk},
    \label{Eq.IV_formation_energy}
\end{equation}
where $N$ denotes the number of atoms in the respective supercell. 

All AFQMC and CCSD(T) calculations in this work are twist-averaged, the HSE calculations are converged with respect to BZ sampling and basis set, and the RPA is evaluated in both setups. We label each result accordingly: AFQMC-ta, CCSD(T)-ta, and RPA-ta denote twist-averaged results, RPA-conv and HSE-conv denote converged results, and the suffix $+\Delta$RPA denotes the correction of Eq.~\eqref{Eq.III.1_embedding}. Method names without a suffix refer to the methods in general.

\subsection{Time-step errors}\label{Sec.IV.1_time-step_errors}
We investigated time-step errors in the AFQMC formation energies of the 16-atom self-interstitial and vacancy cells using four imaginary-time steps $\tau$. We omitted the RPA correction here, since it does not depend on $\tau$.

Typically, the time-step error in the AFQMC total energy follows the relation
\begin{equation}
    E(\tau) = E_0 + \alpha\tau + \beta\tau^2,
    \label{Eq.IV.1_extrapolation_total_energy}
\end{equation}
where the quadratic term emerges directly from the short-time propagator. Per time step, the HS transformation in Eq.~\eqref{Eq.II.1_HS_transformation} introduces an error of $\mathcal{O}(\tau^2)$ for non-commuting operators, and the second-order Trotter--Suzuki decomposition in Eq.~\eqref{Eq.II.1_trotter_suzuki} an error of $\mathcal{O}(\tau^3)$. The linear term is attributed to perturbations introduced to stabilize the MC algorithm, such as the cosine projection or rare-event control~\cite{sukurma2024toward}. For energy differences, however, \citeauthor{sukurma2024toward} reported a more robust extrapolation using
\begin{equation}
    E(\tau) = E_0 + \beta\tau^2.
    \label{Eq.IV.1_extrapolation_energy_difference}
\end{equation}
Our formation energies corroborate this behavior. The linear terms largely, though rarely perfectly, cancel in energy differences. Because formation energies vary on a much smaller energy scale than total energies, a three-parameter fit is highly sensitive to statistical noise. Setting $\alpha = 0$ [see Eq.~\eqref{Eq.IV.1_extrapolation_energy_difference}] therefore yields a considerably more robust extrapolation, at the cost of neglecting any residual linear contribution.

Figure~\ref{Fig.IV.1_time-step_error} shows the time-step dependence of the twist-averaged AFQMC formation energies for the self-interstitials [panels (a) X, (b) H, and (c) T] and the vacancy [panel (d) V], extrapolated using Eq.~\eqref{Eq.IV.1_extrapolation_energy_difference} (dashed lines). For all investigated systems, the formation energies at the two smallest time steps (\SI{2.5e-4}{\per\electronvolt} and \SI{5.0e-4}{\per\electronvolt}) agree with the extrapolated values (open symbols at $\tau=0$) within their statistical uncertainties. 

For the 16-atom cells, extrapolating from $\tau=\SI{1.0e-3}{\per\electronvolt}$, the time step used for the 64-atom cell calculations, to $\tau=\SI{0}{\per\electronvolt}$ lowers the formation energies by \SI{28}{\milli\electronvolt}, \SI{54}{\milli\electronvolt}, \SI{54}{\milli\electronvolt}, and \SI{19}{\milli\electronvolt} for X, H, T, and V, respectively.

\subsection{Finite-size errors}\label{Sec.IV.2_bulk_energies}

\begin{table}
\caption{\label{Tab.IV.2_bulk_energies} Energies per atom (in \si{\electronvolt}) of bulk silicon in the 16- and 64-atom cells. The suffixes -conv and -ta denote converged and twist-averaged energies; the difference between RPA-conv and RPA-ta is the RPA correction $\Delta E^\mathrm{RPA}$ [see Eq.~\eqref{Eq.III.1_rpa_correction}], split into its EXX and correlation parts. CCSD(T)-ta includes the FPC. The AFQMC calculations used a time step of \SI{1.0e-3}{\per\electronvolt}. Statistical uncertainties ($1\sigma$) are given in parentheses.}
\begin{ruledtabular}
\begin{tabular}{l
                S[table-format = -2.3(1)]
                S[table-format = -2.3(1)]}
                        & {16 atoms}   & {64 atoms}   \\
\midrule
RPA-conv                         & -15.365    & -15.364    \\
RPA-ta                           & -15.590    & -15.289    \\
$\Delta E^\mathrm{RPA}$          & 0.226      & -0.076     \\
\quad EXX part                   & 0.837      & 0.189      \\
\quad correlation part           & -0.611     & -0.265     \\
\midrule
AFQMC-ta                         & -13.790(1) & -13.495(1) \\
AFQMC+$\Delta$RPA                & -13.564(1) & -13.571(1) \\
CCSD(T)-ta                       & -13.763    &            \\
\end{tabular}
\end{ruledtabular}
\end{table}

Table~\ref{Tab.IV.2_bulk_energies} lists the energies per atom of bulk silicon in the 16- and 64-atom supercells obtained with converged RPA, twist-averaged RPA, AFQMC, and CCSD(T), as well as AFQMC+$\Delta$RPA. To allow a like-for-like comparison, the AFQMC-ta energies were obtained using a time step of $\tau = \SI{1.0e-3}{\per\electronvolt}$; the 16-atom total energies are not time-step extrapolated. Although the statistical uncertainties per atom are similar for both supercell sizes, the uncertainties of the total energies are four times larger for the 64-atom supercell.

Both the AFQMC-ta and RPA-ta energies per atom differ by approximately \SI{300}{\milli\electronvolt} between the 16- and 64-atom supercells, whereas the RPA-conv energies agree to within \SI{1}{\milli\electronvolt}. Twist-averaged RPA and AFQMC thus show similar finite-size errors, which is the premise of Eq.~\eqref{Eq.III.1_embedding}. With the correction, the AFQMC+$\Delta$RPA energies per atom of the two cells agree to within \SI{7}{\milli\electronvolt}. This agreement suggests that the RPA correction effectively captures the missing long-range exchange and correlation effects even in the 16-atom cell. Since the converged RPA energies of the two cells agree to within \SI{1}{\milli\electronvolt}, the remaining \SI{7}{\milli\electronvolt} reflect the difference between the finite-size errors of the twist-averaged AFQMC and RPA energies. They may stem from two-body finite-size errors in the short-range correlation, which AFQMC in the 16-atom cell does not yet fully capture and the RPA does not correct.

The correction itself changes sign: it raises the energy per atom by \SI{226}{\milli\electronvolt} in the 16-atom cell but lowers it by \SI{76}{\milli\electronvolt} in the 64-atom cell. To trace this sign change, we split the correction into its EXX and correlation contributions (see Table~\ref{Tab.IV.2_bulk_energies}). As expected, the two contributions have opposite signs and partly cancel, but the degree of cancellation can vary with cell size for two reasons: first, the finite-size error of the correlation energy may decay more slowly than that of the exchange energy, and second, the correlation part also contains most of the basis-set correction, which does not depend on the cell size. In addition, $\Delta E^\mathrm{RPA}$ corrects residual one-body finite-size errors due to the finite twist-averaging grid; since the 16- and 64-atom cells use equivalent grids, we expect this contribution to be the same per atom in both cells.

\begin{table*}[ht!]
\caption{\label{Tab.IV.3_formation_energies} Formation energies (in \si{\electronvolt}) of silicon self-interstitials and vacancies in 16- and 64-atom cells from k-point- and basis-set-converged HSE and RPA, twist-averaged RPA, AFQMC, and CCSD(T) calculations, and finite-size (FS) corrected AFQMC+$\Delta$RPA. The 16-atom AFQMC values are extrapolated to $\tau\to0$; the 64-atom values use $\tau=\SI{1.0e-3}{\per\electronvolt}$. Vacancy entries refer to the undistorted vacancy (V) for 16-atom cells and the Jahn--Teller-distorted vacancy (VJT) for 64-atom cells. Statistical uncertainties $(1\sigma)$ are given in parentheses.}
\begin{ruledtabular}
\begin{tabular}{lccccccccccc}
    & \multicolumn{4}{c}{converged} & \multicolumn{5}{c}{twist average} & \multicolumn{2}{c}{FS corrected} \\
    \cmidrule{2-5}\cmidrule{6-10}\cmidrule{11-12}
    & \multicolumn{2}{c}{HSE} & \multicolumn{2}{c}{RPA} & \multicolumn{2}{c}{RPA} & \multicolumn{2}{c}{AFQMC} & CCSD(T) & \multicolumn{2}{c}{AFQMC+$\Delta$RPA} \\
    \cmidrule{2-3}\cmidrule{4-5}\cmidrule{6-7}\cmidrule{8-9}\cmidrule{10-10}\cmidrule{11-12}
supercell  & 16   & 64   & 16   & 64   & 16   & 64   & 16      & 64       & 16   & 16      & 64       \\
\midrule
X         & 4.17 & 4.12 & 4.23 & 4.24 & 4.46 & 4.13 & 4.64(1) & 4.38(11) & 4.48 & 4.41(1) & 4.49(11) \\
H         & 4.35 & 4.21 & 4.42 & 4.34 & 4.64 & 4.28 & 4.79(1) & 4.55(11) & 4.64 & 4.57(1) & 4.62(11) \\
T         & 4.44 & 4.63 & 4.49 & 4.74 & 4.73 & 4.61 & 5.11(2) & 5.60(14) & 4.86 & 4.87(2) & 5.73(14) \\
V/VJT     & 3.67 & 4.69 & 3.49 & 4.44 & 3.51 & 4.20 & 3.81(1) & 4.97(19) & 3.70 & 3.79(1) & 5.20(19) \\
\end{tabular}
\end{ruledtabular}
\end{table*}

The CCSD(T)-ta energy per atom of the 16-atom cell lies \SI{27}{\milli\electronvolt} above the AFQMC-ta value at $\tau = \SI{1.0e-3}{\per\electronvolt}$ and \SI{30}{\milli\electronvolt} above the time-step-extrapolated value. Since both calculations employ identical computational parameters and start from the same HF reference, we suspect that this difference reflects a fundamental difference between the two methods. Most of this difference cancels in formation energies, but not all (see Sec.~\ref{Sec.IV.3_formation_energies}). Whether a full treatment of triple excitations would bring CCSD(T) and AFQMC closer remains an open question that we could not address because of computational constraints.

\subsection{Formation energies}\label{Sec.IV.3_formation_energies}

Table~\ref{Tab.IV.3_formation_energies} summarizes our formation energies of silicon self-interstitials (X, H, T) and vacancies (V, VJT) in 16- and 64-atom cells obtained with HSE, RPA, AFQMC, and CCSD(T) using converged BZ and basis-set as well as twist-average setups. The difference between the converged and twist-averaged RPA columns is the RPA correction applied to AFQMC in the last column [Eqs.~\eqref{Eq.III.1_embedding} and~\eqref{Eq.III.1_rpa_correction}]. In this section, we first discuss the RPA and HSE results and the cell-size dependence, then assess the RPA correction to AFQMC-ta, and finally compare AFQMC-ta with CCSD(T)-ta.

In general, the RPA systematically underestimates the defect formation energies relative to AFQMC and CCSD(T), a trend well documented in the literature~\cite{harl2010assessing,Olsen2013Beyond,kaltak2014cubic}. This is attributed to the RPA's inaccurate description of localized defect states: missing short-range exchange-correlation effects lead to a disproportionate overestimation of the correlation energy in the defect cell compared to the pristine crystal. The underestimation is much larger relative to AFQMC than to CCSD(T): in the 16-atom cells, where RPA-ta, AFQMC-ta, and CCSD(T)-ta share the same twist-average setup, RPA-ta lies \SI{0.02}{\electronvolt} below CCSD(T)-ta for X and agrees with it to within \SI{0.01}{\electronvolt} for H, but lies \SI{0.18}{\electronvolt} and \SI{0.15}{\electronvolt} below AFQMC-ta for X and H, respectively.

At the RPA-conv level, the formation energies of X and H change by at most \SI{80}{\milli\electronvolt} between the 16- and 64-atom cells. In contrast, the formation energies of T and the vacancy increase by \SI{0.25}{\electronvolt} and \SI{0.95}{\electronvolt}, respectively. For the vacancy, the increase occurs even though the Jahn--Teller distortion slightly stabilizes the defect in the 64-atom cell.
This pronounced cell-size dependence is likely an artifact of the small defect cell: for T and the vacancy, the partially occupied defect levels interact with their periodic images and form dispersive, partially filled bands. This spuriously stabilizes the defect (bonding linear combinations are occupied). Defects with fully occupied levels, such as X and H, do not benefit from this band formation and show a much weaker cell-size dependence.
These trends agree with a previous RPA study by \citeauthor{kaltak2014cubic}~\cite{kaltak2014cubic} and show that 16-atom cells are inadequate for realistic formation energies of the T interstitial and the vacancy.

Overall, the HSE hybrid functional performs very similarly to the RPA but predicts formation energies slightly lower than RPA-conv for the interstitials and \SIrange{0.18}{0.25}{\electronvolt} above RPA-conv for the vacancies. The HSE-conv formation energies show only a minor cell-size dependence for X and H, whereas those of T and the vacancy increase by \SI{0.19}{\electronvolt} and \SI{1.02}{\electronvolt}, respectively.

For X and H, the 16-atom AFQMC-ta and RPA-ta formation energies are approximately \SI{300}{\milli\electronvolt} larger than the 64-atom values, while the converged RPA formation energies of these defects change by at most \SI{80}{\milli\electronvolt}. We therefore attribute the twist-average differences mostly to residual two-body finite-size errors in the 16-atom results, \textit{i.e.}, too coarse a sampling of the momentum transfer in the fluctuations. Adding the RPA correction to AFQMC restores excellent agreement between the 16-atom and 64-atom AFQMC+$\Delta$RPA formation energies. Including the time-step errors found for the 16-atom cells (see Sec.~\ref{Sec.IV.1_time-step_errors}) would lower the 64-atom defect formation energies of X and H by \SI{28}{\milli\electronvolt} and \SI{54}{\milli\electronvolt}, tentatively aligning the changes from 16-atom to 64-atom cells even more closely with the RPA-conv results.

\begin{figure*}[ht!]
    \centering
    \includegraphics{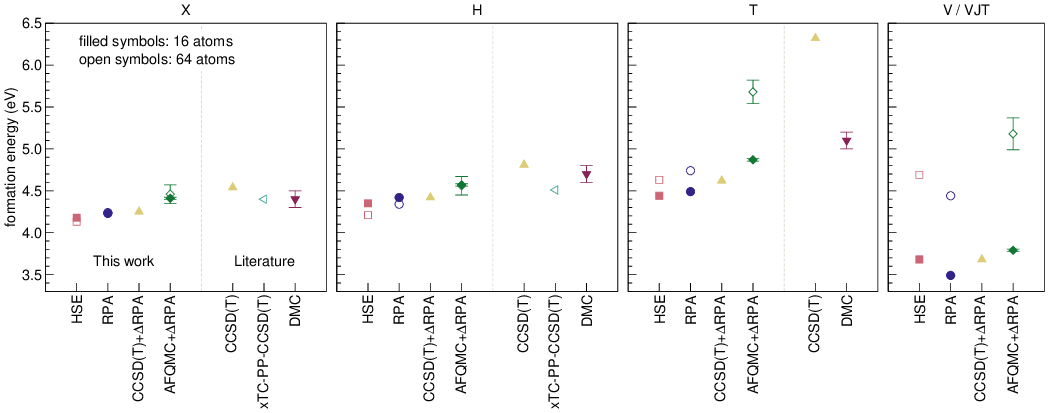}
    \caption{Formation energies (in \si{\electronvolt}) of silicon self-interstitials (X, H, T) and vacancies (V, VJT) in the thermodynamic limit from HSE, RPA, CCSD(T)+$\Delta$RPA, and AFQMC+$\Delta$RPA. Filled symbols: 16-atom cells; open symbols: 64-atom cells. Vacancy values refer to the undistorted structure (V) in 16-atom cells and the Jahn--Teller-distorted structure (VJT) in 64-atom cells. Literature benchmarks from CCSD(T)~\cite{Salihbegovi_Fomration}, xTC-PP-CCSD(T)~\cite{Simula2026Transcorrelated}, and DMC~\cite{Parker2011Accuracy} are shown for comparison. Error bars denote one standard deviation $(1\sigma)$.}
    \label{Fig.IV.3_formation_energies}
\end{figure*}

For T in the 16-atom cell, the RPA correction lowers the AFQMC-ta formation energy by \SI{240}{\milli\electronvolt}. In the 64-atom cell, the RPA correction is comparable to the statistical uncertainty, suggesting that the two-body finite-size errors are small at this cell size.
For the vacancy, the correction is negligible in the 16-atom cell (\SI{20}{\milli\electronvolt}) but amounts to \SI{230}{\milli\electronvolt} in the 64-atom cell, exceeding the statistical uncertainty of \SI{190}{\milli\electronvolt}. This is the only defect where the RPA correction increases with cell size.

In the 16-atom cells, CCSD(T)-ta yields lower formation energies than AFQMC-ta for all defects: by \SI{160}{\milli\electronvolt}, \SI{150}{\milli\electronvolt}, and \SI{110}{\milli\electronvolt} for X, H, and V, respectively, and the largest deviation is observed for T with \SI{250}{\milli\electronvolt}.
We stress that this is the only internally consistent comparison between the two methods: both use identical computational parameters, are twist-averaged over the same k-points, start from the same HF reference, and neither corrects for two-body finite-size errors. Apart from the many-body method itself, the calculations differ only in the treatment of the unoccupied orbitals: AFQMC naturally includes all excitations into the unoccupied manifold, whereas for CCSD(T), we truncated the unoccupied manifold to 20 natural orbitals per occupied state and corrected the resulting basis-set incompleteness error with the FPC. 

Our final AFQMC formation energies in the thermodynamic limit are the AFQMC+$\Delta$RPA values since the RPA correction effectively removes the two-body finite-size errors that remain after twist averaging. We set these values in bold in Table~\ref{Tab.IV.4_literature_comparison}.
For X and H, we recommend the 16-atom values: they agree with the corresponding 64-atom results within statistical uncertainties, are extrapolated to the zero-time-step limit, and have statistical uncertainties an order of magnitude smaller than those of the 64-atom calculations.
For T and the vacancy, we recommend the 64-atom values, because defect--defect interactions of the periodic images in the 16-atom cell spuriously stabilize the defects. Furthermore, only the 64-atom cell hosts the Jahn--Teller-distorted vacancy. These 64-atom values are not time-step extrapolated. As the agreement for X and H indicates, their time-step errors lie below the statistical uncertainties.

\subsection{Comparison with literature}\label{Sec.IV.4_literature_comparison}
Figure~\ref{Fig.IV.3_formation_energies} illustrates the energetic landscape of the studied silicon defects in the thermodynamic limit using different methods: HSE, RPA, AFQMC+$\Delta$RPA, and CCSD(T)+$\Delta$RPA alongside literature results. Table~\ref{Tab.IV.4_literature_comparison} compares our AFQMC+$\Delta$RPA and CCSD(T)+$\Delta$RPA formation energies with literature results. All values are corrected toward the thermodynamic limit, but each in a different way: our RPA and HSE results use converged BZ sampling; we correct twist-averaged AFQMC and CCSD(T) with the RPA; the CCSD(T) calculations of Ref.~\onlinecite{Salihbegovi_Fomration} combine k-point-converged HF energies with twist-averaged correlation energies and a structure-factor-based finite-size correction; the DMC calculations of Ref.~\onlinecite{Parker2011Accuracy} employ a single k-point with DFT- and structure-factor-based correction; and the transcorrelated CC calculations [xTC-PP-CCSD(T)] of Ref.~\onlinecite{Simula2026Transcorrelated} embed their results in 64-atom cells. This comparison therefore tests the finite-size strategies as much as the many-body methods themselves.

\begin{table*}[ht!]
\caption{\label{Tab.IV.4_literature_comparison}
Comparison of our AFQMC+$\Delta$RPA and CCSD(T)+$\Delta$RPA formation energies (in \si{\electronvolt}) with literature values from CCSD(T)~\cite{Salihbegovi_Fomration}, DMC~\cite{Parker2011Accuracy}, and embedded xTC-PP-CCSD(T)~\cite{Simula2026Transcorrelated}. CCSD(T)+$\Delta$RPA denotes our 16-atom CCSD(T) values with the same RPA correction as AFQMC+$\Delta$RPA. The bottom rows list the separations $\Delta E_\mathrm{XH} = E_\mathrm{F}(\mathrm{H}) - E_\mathrm{F}(\mathrm{X})$ and $\Delta E_\mathrm{HT} = E_\mathrm{F}(\mathrm{T}) - E_\mathrm{F}(\mathrm{H})$. Our recommended AFQMC+$\Delta$RPA values are set in bold. Statistical uncertainties $(1\sigma)$ are given in parentheses; for the separations, they are upper bounds, since the contribution of the bulk reference cancels. The 16-atom AFQMC+$\Delta$RPA results are time-step extrapolated; the 64-atom results were obtained using $\tau=\SI{1.0e-3}{\per\electronvolt}$.}
\begin{ruledtabular}
\begin{tabular}{lcccccc}
    & \multicolumn{3}{c}{This work} & \multicolumn{3}{c}{Literature} \\
    \cmidrule{2-4}\cmidrule{5-7}
    & \multicolumn{2}{c}{AFQMC+$\Delta$RPA} & CCSD(T)+$\Delta$RPA & CCSD(T)~\cite{Salihbegovi_Fomration} & DMC~\cite{Parker2011Accuracy} & xTC-PP-CCSD(T)~\cite{Simula2026Transcorrelated} \\
    \cmidrule{2-3}\cmidrule{4-4}\cmidrule{5-5}\cmidrule{6-6}\cmidrule{7-7}
supercell              & 16      & 64       & 16   & 16   & 16     & 64 \\
\midrule
X                      & \textbf{4.41(1)} & 4.49(11) & 4.25 & 4.54 & 4.4(1) & 4.40 \\
H                      & \textbf{4.57(1)} & 4.62(11) & 4.42 & 4.81 & 4.7(1) & 4.51 \\
T                      & 4.87(2) & \textbf{5.73(14)} & 4.62 & 6.32 & 5.1(1) &      \\
V/VJT                  & 3.79(1) & \textbf{5.20(19)} & 3.68 &      &        &      \\
\midrule
$\Delta E_\mathrm{XH}$ & 0.16(1) & 0.13(16) & 0.17 & 0.27 & 0.3(1) & 0.11 \\
$\Delta E_\mathrm{HT}$ & 0.30(2) & 1.11(18) & 0.20 & 1.51 & 0.4(1) &      \\
\end{tabular}
\end{ruledtabular}
\end{table*}

Specifically, we emphasize that the present twist-average CCSD(T) correlation energies match within a few tens of \si{\milli\electronvolt} those calculated previously by \citeauthor{Salihbegovi_Fomration}~\cite{Salihbegovi_Fomration}. \citeauthor{Salihbegovi_Fomration} used a setup very similar to the present work, but sampled the twist-average k-points stochastically. This has only a minor effect on the correlation energies. Hence, in this particular case, the differences in the thermodynamic limit are almost entirely due to different finite-size corrections. 
\citeauthor{Salihbegovi_Fomration} attributed the overestimation of CCSD(T) for T to strong static correlation arising from the threefold-degenerate highest occupied orbital, which hosts two electrons. They suspected that a multi-configurational character of the interstitial challenges single-reference CCSD(T).
Our results point instead to the finite-size treatment. Their workflow combines k-point-converged HF energies with twist-averaged correlation energies and applies a structure-factor-based finite-size correction (Sec.~\ref{Sec.III.3_ccsd(t)_calculations}). Reproducing this workflow with our CCSD(T) data, we obtain formation energies within \SI{0.11}{\electronvolt} of Ref.~\onlinecite{Salihbegovi_Fomration} for X and H, and \SI{6.72}{\electronvolt} for T, which is even larger than the value in Ref.~\onlinecite{Salihbegovi_Fomration} (\SI{6.32}{\electronvolt}).
Combined with twist-averaged HF energies and the RPA correction, the same correlation energies yield \SI{4.62}{\electronvolt} for T [see CCSD(T)+$\Delta$RPA in Table~\ref{Tab.IV.4_literature_comparison}]. Since both values stem from the same single-reference CCSD(T) calculations, multi-configurational character cannot account for their difference of \SI{2.1}{\electronvolt}. However, we cannot exclude a smaller multi-configurational contribution: among the four defects, T shows the largest difference between AFQMC-ta and CCSD(T)-ta (\SI{250}{\milli\electronvolt}).
Overall, we conclude that the previous CCSD(T) calculations used somewhat inaccurate finite-size corrections, and will hence exclude them from further discussion.

We first consider X and H, where the cell-size dependence is small. The main message is that the agreement between AFQMC+$\Delta$RPA and xTC-PP-CCSD(T) is excellent for both the absolute and relative formation energies of the two defects. The energy difference between X and H is slightly larger for AFQMC+$\Delta$RPA with 16 atoms than for xTC-PP-CCSD(T). However, one should recall that the difference decreased from 16 to 64 atoms in the converged RPA calculations, so one would expect a similar behavior for AFQMC. In fact, after correcting for time-step errors based on the 16-atom cells, the 64-atom AFQMC+$\Delta$RPA matches the xTC-PP-CCSD(T) very closely. This agreement is likely partly fortuitous because our AFQMC statistical uncertainties are significantly larger. Nevertheless, xTC-PP-CCSD(T) and AFQMC+$\Delta$RPA agree within statistical uncertainties on 64-atom cells. 
However, the 16-atom CCSD(T)+$\Delta$RPA calculations do not quite align.
One might expect them to be consistent with xTC-PP-CCSD(T), since both are  CCSD(T) calculations only handling basis-set errors differently.  
At this point, we cannot resolve the issue, but we want to suggest a few possible culprits. (i) Changing the CCSD(T) bulk energy by \SI{10}{\milli\electronvolt} per atom would perfectly align the CCSD(T)+$\Delta$RPA data with the AFQMC+$\Delta$RPA calculations for X and H. This highlights the difficulty of determining absolute defect formation energies. Generally, the constant offset in the defect energies between CCSD(T) and AFQMC suggests some size-consistency issue.
(ii) The FPC corrections may be less reliable than the numbers suggest, given that we are extrapolating from 20 virtual natural orbitals to about 100 orbitals per occupied state. (iii) Another possible cause is the singularity correction of the perturbative triples, which formally require integrating a diverging term over all momentum transfers. Although the integral is convergent, approximations are used in practice, such as the Gygi--Baldereschi scheme~\cite{Gygi1986Self} used here.
A more careful test is beyond the scope of the present work.

Extending the analysis to include the T interstitial and the vacancy does not alter our conclusion. Again, AFQMC yields slightly larger formation energies, but with the error increasing slightly for T. However, as already discussed, T and the vacancy are metallic, so CCSD(T) or AFQMC may struggle more with the multi-configurational character. When we extend the analysis to DMC, we observe excellent agreement with the 16-atom AFQMC+$\Delta$RPA, mostly within the DMC satistical uncertainties of \SI{0.1}{\electronvolt}. Again, there are slightly larger differences for T, possibly because a multi-configurational character makes it more challenging to treat with correlated wavefunction methods, and because it is more difficult to extrapolate with respect to k-point sampling. 
Finally, we would like to emphasize again that the results for T and V in 16-atom cells are not particularly relevant, since the cell-size dependence is large. Our results for 64 atoms are currently the only reliable estimates for the formation energies of the vacancy and tetrahedral interstitial.

\section{Conclusions}\label{Sec.V_conclusions}

We computed the formation energies of silicon self-interstitials and vacancies using plane-wave AFQMC within the projector-augmented wave framework. 
Our implementation avoids extrapolation with respect to the number of virtual orbitals by omitting an explicit truncation of the excitation space. 
To make the  AFQMC calculations feasible, we use a twist-averaging procedure to sample the Brillouin zone. While this is computationally convenient, it introduces significant errors since the momentum transfer in the fluctuations is restricted to the $\Gamma$ point and small momentum transfers are not sampled. As in our previous work, we correct for this error by subtracting the RPA total energies, also determined using the twist-averaging procedure, and then adding back the RPA total energy, determined at a converged k-point setting. It is, of course, important to check that this procedure is accurate and reliable. We have systematically studied this in our previous work on lattice constants, but it might not be obvious that the same approach works equally well for energies. 
Three observations support the accuracy of the present calculations.
First, the Si bulk energy is consistent to within \SI{7}{\milli\electronvolt} per atom for the 16- and 64-atom cells. Prior to the correction, however, the difference was a significant \SI{300}{\milli\electronvolt}. 
Second, for all methods we have used so far (documented here for the RPA and the HSE hybrid functional but also true for semilocal functionals), the formation energies of the dumbbell (X) and hexagonal (H) interstitials converge well using 16-atom cells. We demonstrate that, with the RPA corrections included, the 16- and 64-atom results agree within statistical uncertainties for AFQMC. For comparison, the uncorrected values differ by about \SI{250}{\milli\electronvolt}.
Finally, for the self-interstitials, the RPA corrections decrease systematically from 16-atom cells to 64-atom cells. This clearly demonstrates the effectiveness of the current correction scheme and reinforces the accuracy of the RPA in describing long-range fluctuations and correlation effects.

Our final AFQMC values, \SI{4.41}{\electronvolt} for X and \SI{4.57}{\electronvolt} for H, align exceedingly well with those from previously published embedded xTC-PP-CCSD(T) calculations. The value for X agrees with DMC within the statistical uncertainty; the value for H lies \SI{0.13}{\electronvolt} below DMC, slightly outside its uncertainty of \SI{0.1}{\electronvolt}. However, our twist-averaged CCSD(T) results, when corrected using the same RPA workflow, yield somewhat lower formation energies than xTC-PP-CCSD(T), AFQMC, and DMC, approaching previously published RPA values. This certainly deserves further investigation, but we could not envisage any computational tests that would resolve the discrepancy. We suspect that the discrepancy arises from hidden size-consistency errors, from the focal-point correction, or from the perturbative triples, whose formally divergent contributions must be regularized.

Another important finding of this study is that 16-atom cells are insufficient to obtain even approximate results for defects that form interacting and partially occupied defect levels or undergo Jahn--Teller distortion. Errors for both the tetrahedral interstitial (T) and the vacancy (V) are significant in 16-atom cells, sometimes exceeding \SI{1}{\electronvolt}. The defect formation energies of T and V are generally much larger than anticipated, approaching \SI{5.73}{\electronvolt} for T and \SI{5.20}{\electronvolt} for the Jahn--Teller-distorted vacancy.
It would be highly relevant to study these defects using other methods that explicitly include correlation effects. However, this is not straightforward since both defects tend to result in near-metallic bands or small band gaps if they are in the Jahn-- Teller-distorted configuration. This poses a serious challenge to any correlated wavefunction method. 

\section*{Acknowledgments}
This research was funded in whole or in part by the Austrian Science Fund (FWF) Grant No. \href{https://doi.org/10.55776/P33440}{10.55776/P33440}. The computational results presented were largely obtained using the Austrian Scientific Cluster (ASC).



\widowpenalties 0\relax
\bibliography{bibliography.bib}

\end{document}